\documentclass[a4paper,11pt]{article}
\usepackage{pos}
\usepackage{subcaption}
\usepackage{cleveref} 

\def\orcid#1{\kern .08em\href{https://orcid.org/#1}{\includegraphics[keepaspectratio,width=0.7em]{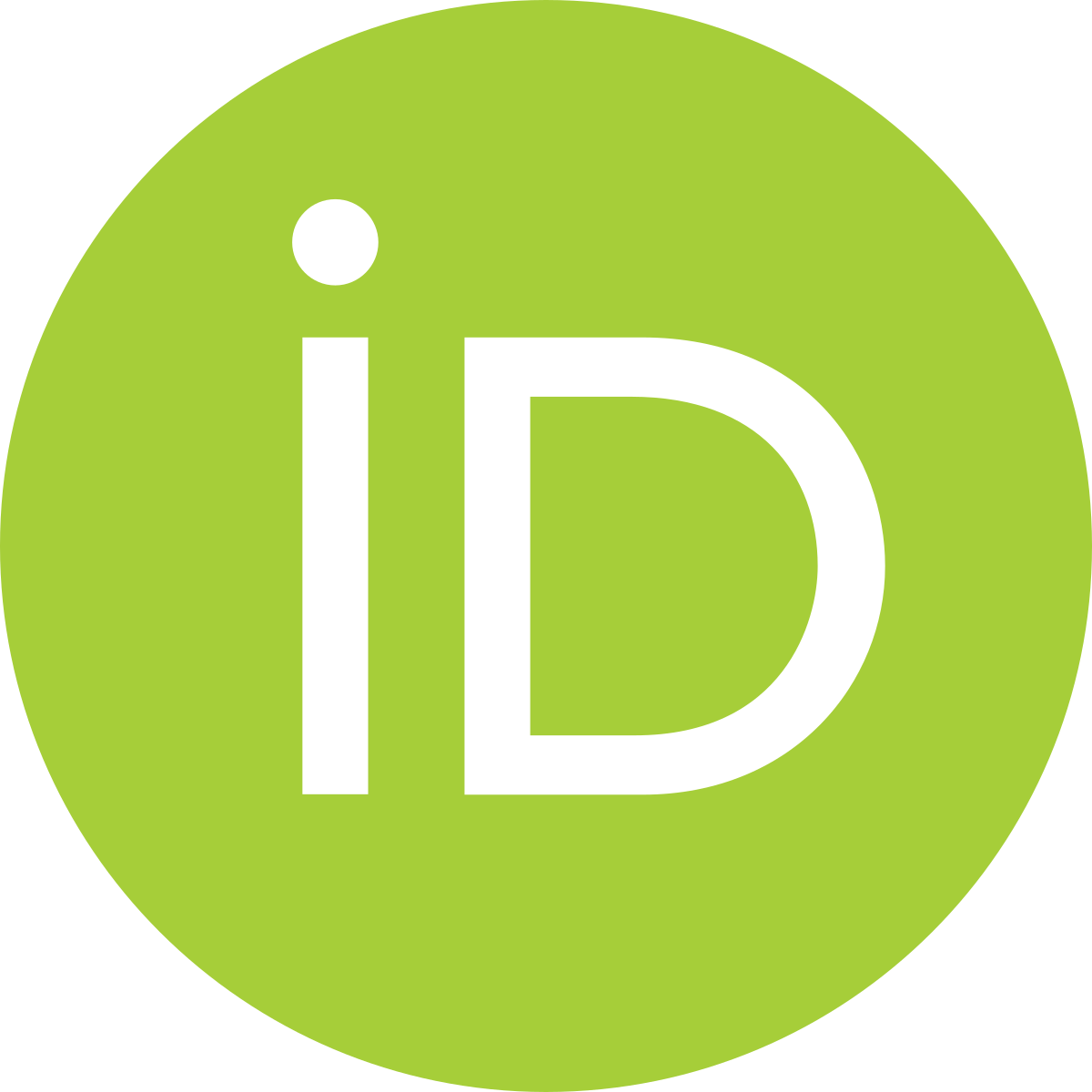}}}

\title{Feature extraction in partial wave analysis using $K$-matrix approach }
\ShortTitle{Feature extraction in partial wave analysis...}

\author*{Adam B. Mapa II\orcid{0009-0003-6301-1567}}
\author{Denny Lane B. Sombillo\orcid{0000-0001-9357-7236}}

\renewcommand{\printHeadAuthors}{AB Mapa II\orcid{0009-0003-6301-1567} and DLB Sombillo\orcid{0000-0001-9357-7236}}

\affiliation{National Institute of Physics, University of the Philippines Diliman,\\
Quezon City 1101, Philippines}

\emailAdd{abmapa@up.edu.ph}
\emailAdd{dbsombillo@up.edu.ph}

\abstract{Structures in the invariant mass distribution are often linked to unstable intermediate states or resonances. In experiments, many signals are detected which have broad, overlapping or intricate profiles, which makes their characterization a formidable task. To ascertain whether or not these enhancements are resonances, and to determine their physical parameters, such as the resonance mass, coupling strength, resonance width, and quantum numbers, a tool known as partial wave analysis (PWA) is employed. To ensure unitarity, and to account for superposing states and non-resonant effects in modeling the scattering amplitude, $K$-matrix parametrization is utilized. To factor in the centrifugal effects on the decay rates arising from breakup processes in nonzero angular momenta, the Blatt-Weisskopf barrier factors are incorporated into the formulation. In this study, $K$-matrix parametrized differential cross sections were generated within near-threshold and resonance energy ranges, which served as the training and validation datasets for the Deep Neural Network (DNN). The Fully-Connected Neural Network (FCNN) architecture is applied to facilitate the classification task of this work, which is the discrimination of dominating partial waves in the scattering processes. The DNN model was designed to encompass various two-hadron scattering phenomena. As a stepping-off point, the values of the thresholds, coupling constants, resonance masses and widths, scattering energy ranges, and the decay channels used in this study found their physical inspiration from pion-nucleon ($\pi$$N$) scattering system in the $S$, $P$, $D$, $F$, and $G$ partial waves. The results of the model's performance showed that the DNN can distinguish the partial wave with resonances from the other partial waves with purely non-resonant contributions, at an accuracy of $69\%$.}

\FullConference{
The 21st International Conference on Hadron Spectroscopy and Structure (HADRON2025)\\
27-31 March, 2025\\
Osaka University, Japan
}

\begin{document}
\maketitle
\section{Introduction}
An array of possible hadronic states have been identified from experiments in recent years. These short-lived signals manifest as peaks in the invariant mass distribution. Investigating the nature of these enhancements and their dynamics has been the central motivation of hadron spectroscopy. To disentangle the ambiguities in the detected signals, partial wave analysis (PWA) \cite{doi:10.1142/S0217751X06034811} is utilized. In practice, solving analytically for the scattering amplitude in QCD is very challenging because of the intrinsically non-perturbative characteristics of the framework in the low-energy domain. In this paper, we propose a hybrid approach to PWA by incorporating a Deep Neural Network (DNN) technique into the conventional parametrization.

\section{\texorpdfstring{$K$}{K}-matrix parametrization}
$K$-matrix formulation provides a parametrization of the scattering amplitude that adheres to unitarity and takes into account overlapping resonances and non-resonant effects. This phenomenological model aims to describe two-body scattering in multichannel or coupled-channel processes. There are different ways of parametrizing $K$-matrix, but the most-widely used form, which is employed in this study, is given by \cite{Chung:1995dx}
\begin{equation}
K_{ij} = \sum_\alpha \frac{g_{\alpha i}(m) g_{\alpha j}(m)}{(m_\alpha^2 - m^2)\sqrt{\rho_i\rho_j}} + a_{ij}+b_{ij}m^2,\label{eq:1}
\end{equation}
where $g_{\alpha i}(m)$ is the residue function which encodes the energy-dependent coupling of the resonance $\alpha$ to channel $i$, $m_\alpha$ is the resonance mass, $\rho_i$ is the phase-space factor in the channel $i$, and $a_{ij}$ and $b_{ij}$ are the non-resonant term coefficients of $K$-matrix. The residue function is defined as
\begin{equation}
g_{\alpha i}(m) = \gamma_{\alpha i} \sqrt{m_\alpha\Gamma_\alpha^0} B_{\alpha i}^l(q, q_\alpha) \sqrt{\rho_i},\label{eq:2}
\end{equation}
where $\gamma_{\alpha i}$ is the coupling strength of the resonance $\alpha$ to channel \(i\), $\Gamma_\alpha^0$ is the resonance width, $B_{\alpha i}^l(q, q_\alpha)$ is the form factor as a ratio of centrifugal barrier factors, $B_{\alpha i}^l(q_i, q_{\alpha i})=F_l(q_i)/F_l(q_{\alpha i})$, with $q_i$ the breakup momentum in channel $i$ and $q_{\alpha i}$ the resonance breakup momentum from resonance $\alpha$ in channel $i$, given angular momentum $l$ \cite{PhysRevD.5.624}.


\section{Differential cross section datasets}
Expressing the transition matrix $T$ in terms of the $K$-matrix, $T=(I-iK\rho)K^{-1}$. The $T$ here is the same quantity as the $T_l(m^2)$ in scattering amplitude definition
\begin{equation}
T(\theta,m^2) = \sum_l (2l + 1) T_l(m^2)P_l(\cos\theta),\label{eq:3}
\end{equation}
where $m$ is the invariant mass and $P_l(\cos\theta)$ is the Legendre polynomial. The resulting form of the differential cross section, which this study utilized for the dataset generation, is
\begin{equation}
\frac{d\sigma}{d\Omega}=\frac{4}{m^2}\frac{\rho_f}{\rho_i}|T(\theta,m^2)|^2.\label{eq:4}
\end{equation}
To generate differential cross sections $d\sigma/d\Omega$ for the DNN training and validation, the physical parameters $m_\alpha$, $\gamma_{\alpha i}$, $\Gamma_\alpha^0$, $a_{ij}$ and $b_{ij}$ are systematically varied within $\pi N$ scattering experiment- or literature-based ranges for each $T$-matrix partial wave amplitude $T_l(m^2)$. To generate the datasets corresponding to the different labels, the resonant term of the $K$-matrix in \eqref{eq:1} is systematically activated depending on which $T_l(m^2)$ the resonances are, while keeping the resonant terms of the remaining $T_l(m^2)$'s inactive, with their non-resonant terms $a_{ij}$ and $b_{ij}$ active. To ensure representative 
coverage of the concerned parameter and functional spaces, the datasets are created by randomly sampling $d\sigma/d\Omega$ over the parameter ranges, cosines of angles $\cos\theta \in [-1,1]$, and energies $E \in [1100, 2800]$ (in MeV). The plots of the $d\sigma/d\Omega$ datasets are shown below.
\begin{figure}[th]
  \centering
  \subfloat[]{\includegraphics[width=0.302\textwidth]{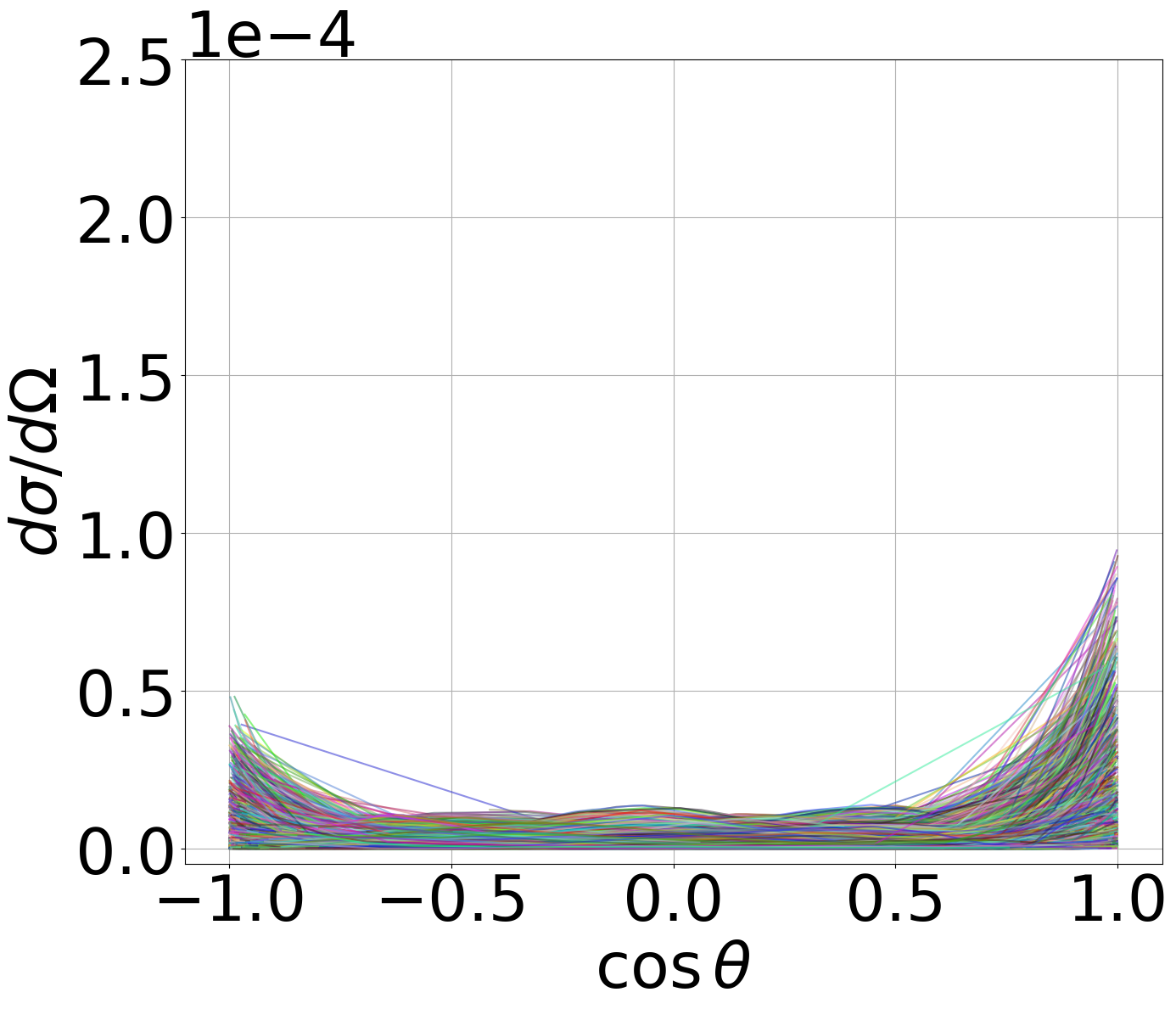}\label{fig:label0}}
  \quad
  \subfloat[]{\includegraphics[width=0.302\textwidth]{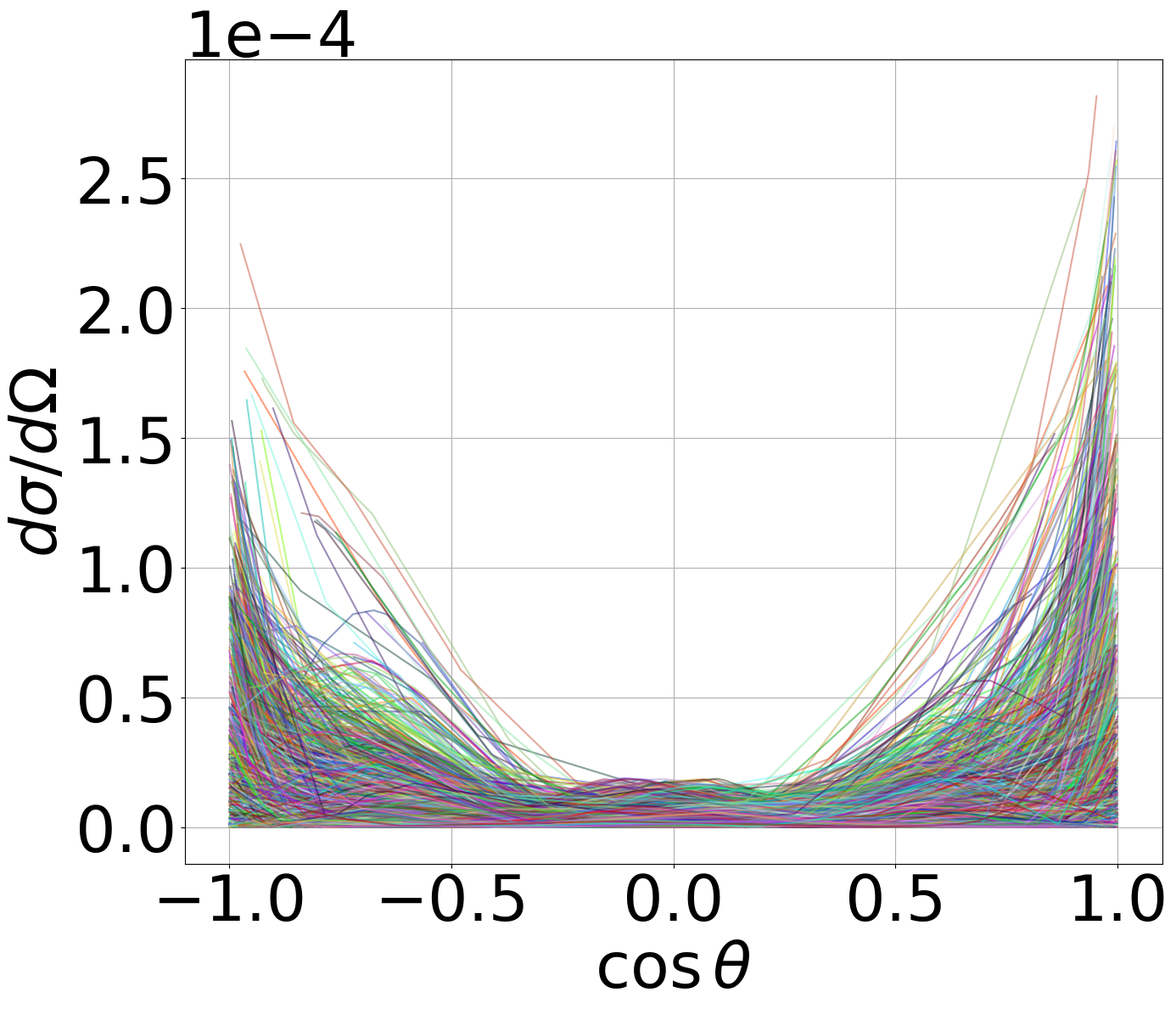}\label{fig:label1}}
  \quad
  \subfloat[]{\includegraphics[width=0.302\textwidth]{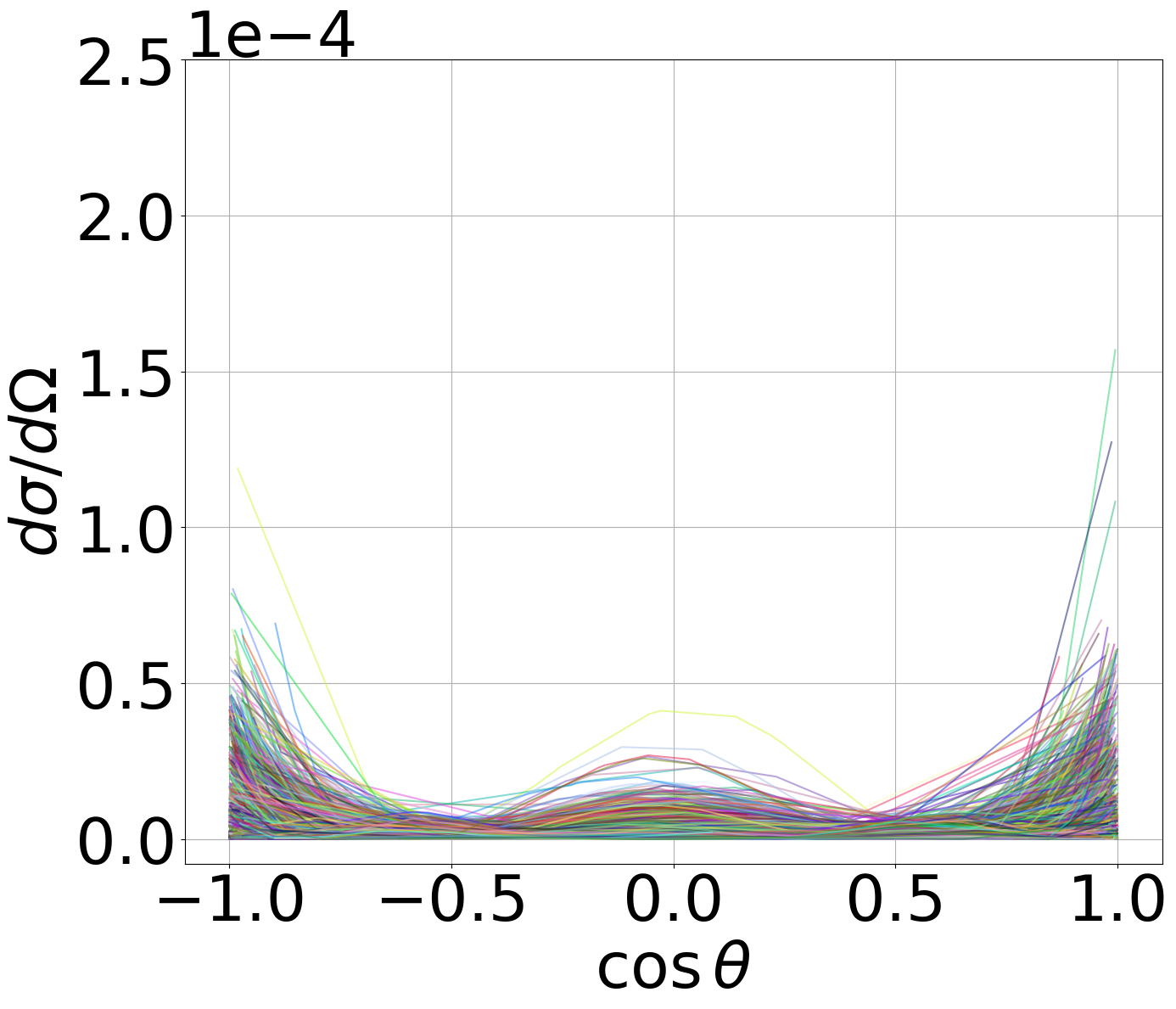}\label{fig:label2}}
  \caption{Differential cross section datasets for the first three labels, each with 10,000 plots of 15 data points. Resonances in partial waves $l=0$, $1$, and $2$ are shown in (a)–(c), respectively.}
  \label{fig:dataset}
\end{figure}

If there were only one partial wave present, then the analysis would be simplified. However, in more realistic scattering processes, there is superposition of contributions from different partial waves. These physical tendencies are mimicked using synthetic datasets as shown in Fig.~\ref{fig:dataset}. The observed characteristic plot behaviors are from instances when the scattering energies are around the resonance masses, or when the non-resonant contributions combine arithmetically to produce constructive effects. In Fig.~\ref{fig:label1}, parabolic behaviors are seen, which come from $|T_{l=1}P_{l=1}|^2$. At the same time, one-bump features in the center manifest, which are from $|T_{l=2}P_{l=2}|^2$ or $|T_{l=4}P_{l=4}|^2$. In Fig.~\ref{fig:label2}, that one-bump behaviors are observed, but the plots also show additional features like the two bumps on the side, which can be traced to $|T_{l=3}P_{l=3}|^2$ or $|T_{l=4}P_{l=4}|^2$. Moreover, there is the smoothening behaviors on the base parts of the plots, which come from the averaging effects of the combined contributions from across all the involved partial waves or, mostly, from $|T_{l=0}P_{l=0}|^2$ in view of Fig.~\ref{fig:label0}. Because of this simultaneous  inter-partial-wave (resonant $+$ non-resonant) mixing, qualitative inferences are suspended, which motivates the DNN approach of this study.

\section{FCNN architecture and performance}
In this paper, a Fully-Connected Neural Network (FCNN) architecture is implemented for the classification task. The FCNN consists of four layers: an input layer [30 neurons = 15 $\cos\theta$ \& 15 $d\sigma/d\Omega$ values], three hidden layers [BatchNorm $+$ ReLU $+$ Dropouts (0.2, 0.2, 0.1)], and an output layer [5 neurons = partial waves (l=0 to l=4)].
\begin{figure}[th!]
\centering
  \makebox[1.0\linewidth]{
  \subfloat[]{\includegraphics[width=0.325\linewidth]{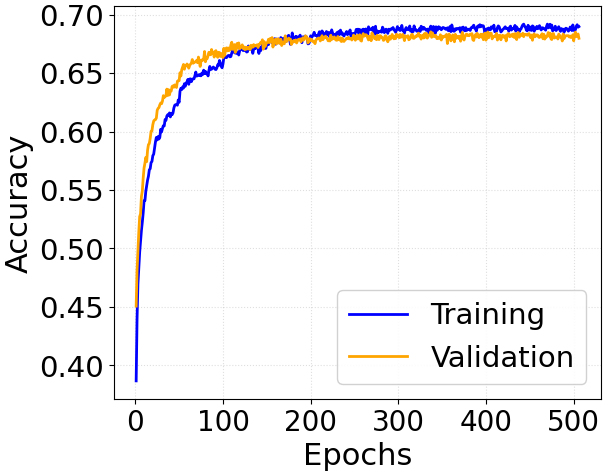}\label{fig:Accuracy}}
  \quad
  \subfloat[]{\includegraphics[width=0.315\linewidth]{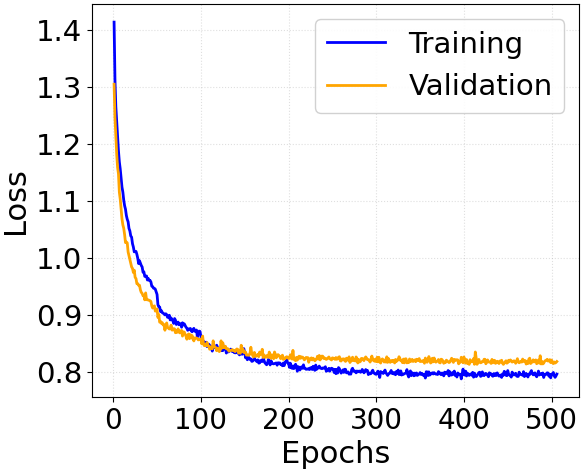}\label{fig:Loss}}
  \quad
  \subfloat[]{\includegraphics[width=0.34\linewidth]{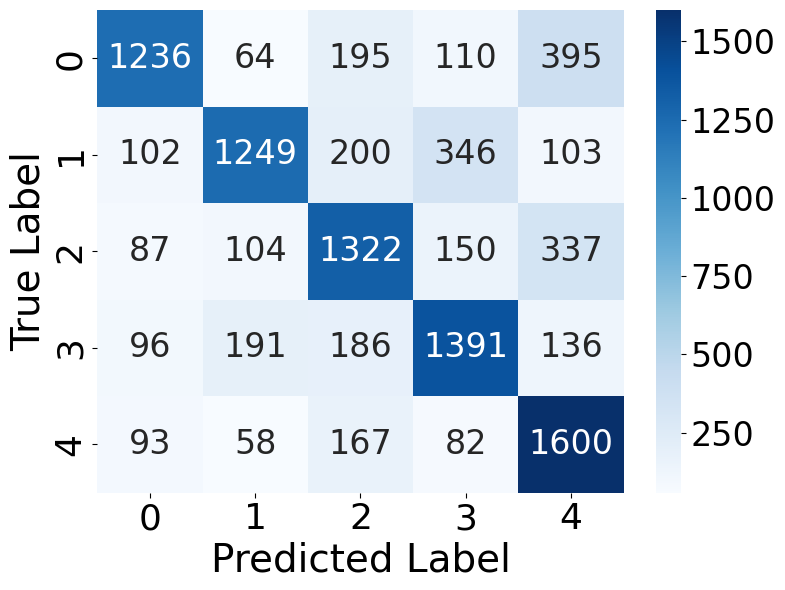}\label{fig:Confusion}}
  }
  \caption{(a) Accuracy plot, (b) loss plot, and (c) confusion matrix illustrating the model’s performance.}
  \label{fig:ModelPerformance}
\end{figure}
In Fig.~\ref{fig:Accuracy} the accuracy increases rapidly in the first 100 epochs and then stabilizes. The training and validation accuracies remain close around 68-70\%, showing that the model is learning reasonably. Furthermore, the model is not overfitting significantly, since training and validation accuracies in Fig.~\ref{fig:Accuracy} and losses in Fig.~\ref{fig:Loss} have close values. This is achieved by using the cross-entropy loss, as this is a multi-class classification problem. The term 'class' refers to either the true label or the predicted label, with 'label' to denote the specific partial wave $l$ with resonance (see Fig.~\ref{fig:dataset}). Figure~\ref{fig:Confusion} shows that the highest number of correctly classified samples is in class 4, with 1600 correct out of 2000. Class 0, with 1236 correctly classified instances, has a notable number of misclassifications into class 4, with 395 cases, which suggests confusion. Classes 1 and 2 have significant misclassifications across multiple classes, particularly into class 3 and class 4, in 346 and 337 instances, respectively. The model also struggles to distinguish 2 and 0, 1, 3, and 4, scoring misclassification values 195, 200, 186, 167, respectively. These off-diagonal values indicate confusion between similarly-behaving classes, as expected based on the generated plots in Fig.~\ref{fig:dataset} with overlapping features, namely the left and right edge parts and the congested base parts of the plots, and other subtle concurrent qualities.

\section{Conclusion and future prospects}
With $K$-matrix parametrization, generated differential cross section datasets are trained and tested using the FCNN architecture. Accuracy values indicated the model learned fairly well at around 69\%. The small gaps between losses may suggest some generalization issue, but is not severe. The next stage of this study will take into account the energy distribution or total cross section. We also hope to improve the accuracy using the Convolutional Neural Network (CNN).

\acknowledgments
ABMII acknowledges the scholarship support provided by the DOST-ASTHRDP.
\bibliographystyle{JHEP}
\bibliography{bib_Mapa}

\end{document}